\documentclass[aip,graphicx]{revtex4-1}

\usepackage{graphicx}
\usepackage{epstopdf}
\usepackage{xcolor}

\usepackage{amssymb}
\usepackage[centertags]{amsmath}
\usepackage{amsthm}
\usepackage[mathcal]{euscript}
\usepackage{tabularx}
\usepackage{fancybox}
\usepackage{wrapfig}

\newcommand{\pD}[2]{\frac{\partial #1}{\partial #2}}
\newcommand{\dD}[2]{\frac{d #1}{d #2}}

\begin{document}


\title{Viscous effects on the Rayleigh-Taylor instability with background temperature gradient} 



\author{S. Gerashchenko}
 \affiliation{ 
MPA-CMMS/CNLS, Los Alamos National Laboratory, Los Alamos, New Mexico 87545, USA
}%

\author{D. Livescu}%
\email{livescu@lanl.gov.}
\affiliation{ 
CCS-2, Los Alamos National Laboratory, Los Alamos, New Mexico 87545, USA
}%


\begin{abstract}

The growth rate of the compressible Rayleigh-Taylor instability is studied in the presence of a background temperature gradient, $\Theta$, using a normal mode analysis. The effect of $\Theta$ variation is examined for three interface types corresponding to combinations of the viscous properties of the fluids (inviscid-inviscid, viscous-viscous and viscous-inviscid) at different Atwood numbers, $At$, and, when at least one of the fluids' viscosity is non-zero, as a function of the Grashof number. For the general case, the resulting ordinary differential equations are solved numerically; however, dispersion relations for the growth rate are presented for several limiting cases. An analytical solution is found for the inviscid-inviscid interface and the corresponding dispersion equation for the growth rate is obtained in the limit of a large $\Theta$. For the viscous-inviscid case, a dispersion relation is derived in the incompressible limit and $\Theta=0$. Compared to $\Theta=0$ case, the role of $\Theta<0$ (hotter light fluid) is destabilizing and becomes stabilizing when $\Theta>0$ (colder light fluid). The most pronounced effect when $\Theta\neq 0$ is found at low $At$ and/or at large perturbation wavelengths relative to the domain size for all interface types. On the other hand, at small perturbation wavelengths relative to the domain size, the growth rate for $\Theta<0$ case exceeds the infinite domain incompressible constant density result. The results are applied to two practical examples, using sets of parameters relevant to Inertial Confinement Fusion coasting stage and solar corona plumes. The role of viscosity on the growth rate reduction is discussed together with highlighting the range of wavenumbers most affected by viscosity. The viscous effects further increase in the presence of background temperature gradient, when the viscosity is temperature dependent.
\end{abstract}


\maketitle

\section{Introduction}
\label{sec:1}

The Rayleigh-Taylor instability \cite{Rayleigh1900, Taylor1950, Sharp1984,Livescu13} (RTI)   occurs in a number of important natural phenomena and applications, for example in supernova explosions and neutron stars, \cite{jun1995, Cabot2006, Litwin2001} solar corona, \cite{Diaz2012, Khomenko2014} earth oceans, atmosphere and mantle, \cite{Davey1981, Schultz2006, Neyret1997, Gerya2003, Mege2000, kelley1976} quantum plasma, \cite{bychkov2008, cao2008, wang2012, momeni2013} combustion, \cite{beale1999} Inertial Confinement Fusion (ICF). \cite{Lindl1998, Atzeni2004} Compared to its classical formulation, in most practical cases, RTI manifests itself as an extremely complex process. The complexity arises, in particular, due to inter-wined manifold of  factors  involved, among which are the density difference,  compressibility, temperature distribution, viscosity, surface tension and other interfacial phenomena for the immiscible case or mass diffusion for the miscible case, heat  diffusion, geometrical and finite boundary effects, specific plasma and magnetic field properties, etc. A lot of endeavor has been undertaken to understand the implication of these  parameters and their combinations for RTI growth. The stabilizing effects of viscosity, surface tension and magnetic fields on the linear stage development were discussed in the classical work of Chandrasekhar. \cite{Chandrasekhar1981} Inclusion of mass diffusion was shown to dump to zero the instability growth rate in the limit of large wave numbers. \cite{Duff1962,LCG07,WL12}. The parameter space increases substantially for the compressible case, since various aspects (e.g. flow compressibility, material properties such as specific heat ratio or viscosity dependence on temperature, background state) depend on different parameters, which independently affect the growth. \cite{Livescu2004,Yu2008,Reckinger2016} Studying the astrophysical phenomena and ICF has inspired further interest in understanding the RTI development for the compressible case, sometimes in association with other phenomena such as plasma effects, ablation, etc. Specifically for the ICF plasma, the crucial role of the ablative \cite{Takabe1985, Betti1998, Betti2001, Atzeni2004, Bychkov2015} and viscous \cite{Weber2014, Haines2014} effects on RTI was highlighted. 

Temperature differences are often present  across the Rayleigh-Taylor (RT) unstable layers and can modify the instability growth compared to layers of constant temperature. For example, in the solar corona prominences,  the temperature difference can reach  $10^5 K$ and during the ICF coasting or deceleration stage up to $10^7 K$. In oceans and atmosphere, due to presence of inverted temperature regions, denser gas and water can occasionally be dumped over less-dense material. In some specific theoretical studies of liquid-vapor interfaces with temperature differences, the effect of mass and heat transfer was shown to be stabilizing or destabilizing depending on whether the gas phase is hotter or colder than the liquid. \cite{Hsieh1978} In the experimental work of Burgess et al., \cite{Burgess2001} heating from below was applied to RT unstable liquid-gas interface. The authors demonstrate that the restoring force provided by the temperature-dependent surface tension can stabilize the interface. Ho \cite{ho1980} studied RTI with two viscous fluids of equal kinematical viscosities in the presence of heat and mass transfer. Thermal effects on linear and nonlinear RTI in the presence of mass, heat transfer and magnetic field were also studied in Ref. \cite{Allah2000} However, there has been no systematic study of the effects of temperature differences on RTI for a range of parameters such as density difference, viscosity, and background stratification for the compressible case. Previously,  Livescu \cite{Livescu2004} investigated the effects of compressibility on RTI with uniform background temperature and Yu and Livescu \cite{Yu2008} extended the study to cylindrical geometry. This paper addresses a more general case,  in which the background temperature varies linearly in the direction perpendicular to the interface. 
For non-zero temperature conduction coefficient, more complex variations of the background 
temperature would lead to unsteady background states, which prevent the separation of variables 
and the reduction of the governing equations to ordinary differential equations required by the normal mode approach.  

Most studies to date of the linear stage of RTI  have been performed for
inviscid fluids, when additional effects compared to the classical incompressible case were considered. Some examples with viscous fluids can be found in Refs. \cite{Livescu2004,LCG07} and references therein. However, even when viscous effects were considered, the viscosities of the two fluids were commensurate. Nevertheless, in applications such as ICF, RTI can develop between fluids with vastly different viscosities. In the ICF context, this is  due to the viscosity variation with temperature and the large temperature difference between the hot spot and the surrounding material. Thus, the limiting case in which one of the fluids is viscous and the other is inviscid, is practically important. The stability of the viscous-inviscid interfaces has been studied for mixing layers \cite{Hogan1985, Power1990}, but, to our knowledge, not for RTI. Thus, in this study, we present the first investigation of the viscous-inviscid interface in the context of RTI; the growth rate is obtained numerically for the general case, while a dispersion relation is presented for the incompressible case. All derivations and results apply to the case when the specific heats of the two fluids are equal and the only incompressible limit considered is that of infinite interfacial pressure (incompressible flow limit). The incompressible fluid limit (ratio of specific heats, $\gamma\rightarrow \infty$) \cite{Livescu2004,Yu2008,Livescu13} and the effects of different specific heats are not addressed since they are not directly relevant to the applications discussed here.

The paper is organized as follows. Section \ref{sec:2} presents the governing equations and the corresponding zeroth (subsection \ref{sec:zero}) and first order  (subsection \ref{sec:first}) equations subsequent to linearization. Sections \ref{sec:3}, \ref{sec:4} and \ref{sec:7} describe the application of these equations to the inviscid-inviscid, viscous-viscous and viscous-inviscid cases, respectively. Although this is not the first study of the viscous-inviscid interface, \cite{Hogan1985, Power1990} to our knowledge this is the first time when this is applied to RTI and the first time when a dispersion equation for the growth rate is presented. For the inviscid-inviscid interface in the limit of large background temperature gradient, an analytical dispersion equation is derived, which gives a good estimate of the growth rate even for small temperatures in the limit of small Atwood, $At$, and/or for large wave numbers. Section \ref{sec:8} presents a discussion of the results for three different $At$ values at several Grashof, $Gr$, numbers. Estimation values of the dimensional growth rates are numerically obtained in section \ref{sec:ICF}  for the parameters relevant to the ICF coasting stage and solar corona plumes and compared to the existing results in the literature. The effects of temperature dependent viscosity are also considered here. Finally, conclusions are provided in section \ref{sec:6}. The terms in the equations for the general viscous-viscous case are provided in the Appendix.

\section{Governing equations}
\label{sec:2}

Taking the case of two superimposed fluids with an interface at $\hat{x}_1 =
0$ and the gravitational acceleration given by $(-\hat{g},0,0)$, the
equations of motion for each fluid are \cite{RLV10,Livescu13,Reckinger2016}
\begin{equation}\label{eq:1}
\pD{\hat{\rho}}{\hat{t}} + \pD{\hat{\rho}\hat{u}_k}{\hat{x}_k} = 0\:,
\end{equation}
\begin{equation}\label{eq:2}
\pD{\hat{\rho}\hat{u}_i}{\hat{t}} +
\pD{\hat{\rho}\hat{u}_i\hat{u}_k}{\hat{x}_k} = - \pD{\hat{p}}{\hat{x}_i}
+  \pD{\hat{\tau}_{ik}}{\hat{x}_k} -\hat{\rho}\hat{g}\delta_{i1}\:,
\end{equation}
\begin{equation}\label{eq:3}
\pD{\hat{\rho}\hat{e}}{\hat{t}} + \pD{\hat{\rho}\hat{e}\hat{u}_k}{\hat{x}_k} =
-\hat{p}\pD{\hat{u}_k}{\hat{x}_k}+\hat{\tau}_{jk}\pD{\hat{u}_j}{\hat{x}_k}+\pD{}{\hat{x}_k}\left(\hat{\kappa}
  \pD{\hat{T}}{\hat{x}_k}\right)\:,
\end{equation}
where the viscous stress is Newtonian, $\hat{\tau}_{ij} = \hat{\mu}(\partial{\hat{u}_i}/\partial{\hat{x}_j}+\partial{\hat{u}_j}/\partial{\hat{x}_i}-(2/3)(\partial{\hat{u}_k}/\partial{\hat{x}_k})\delta_{ij})$, and the heat flux is assumed to follow Fourier's law. In these equations, a dimensional quantity is denoted by a hat,
$(\hat{\cdot})$.
To close the governing equations, ideal gas equations of state for pressure
and internal energy are used:
\begin{equation}\label{eq:4}
  \hat{T} = \frac{\hat{p}}{\hat{R}\hat{\rho}}\:,  \  \  \hat{e}=\frac{\hat{p}}{\hat{\rho}(\gamma-1)}\:,
\end{equation}
where $\hat{R}$ is the gas constant and $\gamma$ is the ratio of specific heats. In the above equations, material properties such as $\gamma$ and $\hat{\mu}$ can be different for the two fluids, but are constant for each of the fluids unless otherwise specifically considered (e.g. when studying the influence of a temperature dependence of  $\hat{\mu}$).

\noindent
By defining:
\begin{alignat*}{3}
\hat{x}_i &= x_i\hat{L} 
& \qquad \hat{t} &= t\sqrt{\frac{\hat{L}}{\hat{g}}}
& \qquad   \hat{\rho}_m &= {\rho_m} (\hat{\rho}_{1,\infty}+\hat{\rho}_{2,\infty}) 
\\ \hat{u}_{im} &=
{u}_{im}\sqrt{\hat{g}\hat{L}} & \qquad
\hat{p}_m &={p}_m\hat{p}_{\infty} & \qquad \hat{T}_m &= {T}_m\hat{T}_\infty \:,
\end{alignat*}
Eq.(\ref{eq:1}) - (\ref{eq:3}) can be cast  into non-dimensional form. In these definitions, the dimensional quantities defined at the fluid interface are denoted by an ($_\infty$) subscript, ($_m$) subscript indicates a quantity in either fluid 1 or 2, with fluid 2 at the top, non-dimensional quantities are unadorned, and $\hat{L}$ is the height of the domain occupied by each fluid (half-height of the total domain). These 
non-dimensionalizations imply that there is a density jump across the interface, but the pressure and temperature are continuous across the interface. Continuities of the background pressure and temperature assume that the unperturbed configuration is in thermodynamical equilibrium at the interface and are necessary to reduce the first order equations to ordinary differential equations. This implies that the two fluids had been in contact for sufficient time before the perturbation is applied. 

Using the definitions above, the non-dimensional forms of Eq.(\ref{eq:1}) -(\ref{eq:3}) become
\begin{equation}\label{eq:7}
\pD{{\rho}}{{t}} +  \pD{{\rho} {u}_k}{{x}_k} = 0\:,
\end{equation}
\begin{equation}\label{eq:8}
\pD{{\rho}{u}_i}{{t}} +
\pD{{\rho}{u}_i{u}_k}{{x}_k} =
-\frac{1}{M^2}\pD{{p}}{{x}_i} + \frac{\partial}{\partial{x_k}}\left[\frac{\mu}{\sqrt{Gr}}\left(\pD{u_i}{x_k}+\pD{u_k}{x_i}-\frac{2}{3}\pD{u_l}{x_l}\delta_{ik}\right)\right] - {\rho}\delta_{i1}\:,
\end{equation}
\begin{equation}\label{eq:9}
 \pD{{p}}{{t}} +{u}_k\pD{p}{{x}_k}= -\gamma {p}
 \pD{{u}_k}{{x}_k} +  (\gamma-1)\frac{\mu M^2}{\sqrt{Gr}}\left(\pD{u_j}{x_k}+\pD{u_k}{x_j}-\frac{2}{3}\pD{u_l}{x_l}\delta_{jk}\right)\pD{u_j}{x_k} +
 \gamma \pD{}{{x}_k}\left(\frac{\kappa}{Pr\sqrt{Gr}} \pD{{T}}{{x}_k}\right)\:,
\end{equation}

\noindent
The non-dimensional numbers in equations (\ref{eq:7})-(\ref{eq:9}) are: (a) the gravitational Mach number, characterizing the compressibility effects as the ratio between free fall velocity over the distance $\hat{L}$ and isothermal sound speed,  
\begin{equation}\label{eq:10}
M^2 = \hat{g}\hat{L}\left(\frac{\hat{\rho}_{1,\infty} +
    \hat{\rho}_{2,\infty}}{\hat{p}_{\infty}}\right)\:,
\end{equation}

\noindent
(b) the Grashof number, characterizing the importance of buoyancy relative to viscous forces,
\begin{equation}\label{eq:11}
Gr = \frac{\hat{g}\hat{L}^3}{\hat{\nu}_\infty^2}\:,
\end{equation}

\noindent
and (c) the Prandtl number, characterizing the importance of momentum diffusivity relative to thermal diffusivity
\begin{equation}\label{eq:12}
 Pr = \frac{\hat{\nu}_\infty\gamma\ \hat{p}_\infty}{\hat{\kappa}_\infty(\gamma-1)\hat{T}_\infty}\:.
\end{equation}

\noindent
Note that in the context of RTI, the Archimedes number might be used to replace $Gr$. In the non-dimensionalization used here, the Froude number, $Fr$, does not appear explicitly. Nevertheless, since the velocity perturbation amplitude is small, the linearized analysis corresponds to the limit $Fr \rightarrow 0$. On the other hand, neglecting the nonlinear terms in the momentum equations, but keeping the viscous terms, corresponds to the assumption that the Reynolds number is small. However, since $Gr=Re/Fr$ there is no restriction on its values for the linear analysis, so both limits $Gr\rightarrow 0$ and $Gr\rightarrow \infty$ are valid in this context. The nondimensional dynamic viscosity coefficient is defined as
$\mu_m = \frac{ \hat{\mu}_m } { \hat{\nu}_\infty (\hat{\rho}_{1,\infty} + \hat{\rho}_{2,\infty} )}$,
where the kinematic viscosity at the interface is $\nu_\infty\equiv(\hat{\mu}_{1,\infty}/\hat{\rho}_{1,\infty}+\hat{\mu}_{2,\infty}/\hat{\rho}_{2,\infty})/2 $, and the nondimensional conduction coefficient is $\kappa_m = \frac{\hat{\kappa}_m } { \hat{\kappa}_\infty}$,
with $\hat{\kappa}_\infty\equiv (\hat{\kappa}_1+ \hat{\kappa}_2)/2$. With the above notations, $M^2$ and $Gr$ can be independently varied by changing the pressure, $\hat{p}_\infty$, and kinematic  viscosity, $\hat{\nu}_\infty$, at the interface.

The equation of state, Eq.(\ref{eq:4}), is written in non-dimensional form as
\begin{equation}\label{eq:14}
{p}_m = \frac{\rho_m T_m}{\alpha_m}\:,
\end{equation}
where
\begin{equation}\label{eq:15}
  \alpha_m = \frac{\hat{\rho}_{m,\infty}}{\hat{\rho}_{1,\infty} + \hat{\rho}_{2,\infty}}\:.
\end{equation}

The linearized analysis can be performed in two ways. In the classical approach \cite{Chandrasekhar1981}, the linearized equations are assumed valid throughout the domain, in which case the subscript ($_m$) does not appear in equations (\ref{eq:7})-(\ref{eq:8}) and the variables are considered in the sense of generalized functions. The discontinuity at the interface is treated by integrating the vertical momentum equation over a small volume across the interface, which yields a jump condition across the interface. This is the approach followed in this paper. However, for the rest of the derivations, each fluid region is treated separately, in which case the subscript ($_m$) will be used to distinguish between the two fluid regions. Alternately, one can consider the governing equations separately in each fluid region and treat the interface as a boundary. The vertical momentum equations are integrated over each domain separately and the continuity of the normal stress at the interface yields a condition equivalent to the jump condition from the first approach. The two approaches are fully equivalent for the linearized equations.

\subsection{Zeroth-order equations}
\label{sec:zero}

The two fluids are assumed to be initially at rest and the primary variables are written as small perturbations about the equilibrium (background) state, denoted by the subscript ($_0$). For the (unperturbed) equilibrium state, $\bold{u}_0=0$, variables depend on $x_1$ only and the governing equations in each fluid region are
\begin{equation}\label{eq:16}
\pD{\rho_{0m}}{t} = 0\:,
\end{equation}
\begin{equation}\label{eq:17}
  \pD{p_{0m}}{x_1}= -M^2 \rho_{0m}\:,\  \  \  \pD{p_{0m}}{x_2}=  \pD{p_{0m}}{x_3} = 0\:,
\end{equation}
\begin{equation}\label{eq:19}
 \pD{{p_{0m}}}{{t}} = 
 \gamma_m\pD{}{{x}_k}\left( \frac{\kappa_m}{Pr\sqrt{Gr}}
   \pD{{T_{0m}}}{{x}_k}\right)\:.
\end{equation}

As far as we know, all previous studies of the linear stage of compressible RTI neglect the heat conduction term in equation (\ref{eq:19}). Indeed, if the heat conduction term is non-zero, then the background pressure is not constant in time, which prevents the normal mode analysis. Previous studies were able to neglect this term by considering a uniform background temperature. Nevertheless, the heat conduction term also becomes zero for a constant background temperature gradient, provided that the heat conduction coefficient is constant for each fluid. Since in many practical applications such as ICF or astrophysics, RTI occurs in the presence of background temperature variation, here we consider, for the first time, the role of a background temperature gradient. 

The condition that the heat fluxes are equal on both sides of the interface is imposed by
$\hat{\kappa}_1 \pD{{\hat{T}_{01}}}{{\hat{x}}_k} = \hat{\kappa}_2 \pD{{\hat{T}_{02}}}{{\hat{x}}_k}$.
Thus, in dimensional form, it is assumed that the background temperature varies as $\hat{T}_{0m} = \hat{a} \frac{2\hat{\kappa}_m}{\hat{\kappa}_1+\hat{\kappa}_2}\hat{x}_1 + \hat{T}_\infty$, 
with $\hat{a}\equiv 0.5(\hat{T}_2- \hat{T}_1)/\hat{L}$, where $\hat{T}_{1,2}$ are temperatures at $\hat{x}_1=-L$ and  $\hat{x}_2=+L$, respectively, and $\hat{T}_\infty=(\hat{T}_1\hat{\kappa_1}+ \hat{T}_2\hat{\kappa_2})/(\hat{\kappa_1}+\hat{\kappa_2})$ is the temperature at the interface. In non-dimensional form, the temperature variation is

\begin{equation}\label{eq:21}
{T}_{0m} =\Theta \kappa_m M^2 x_1 + 1,
\end{equation} 
where   $\Theta \equiv (\hat{T}_2- \hat{T}_1)/[2\hat{g}\hat{L}(1/\hat{R}_1+1/\hat{R}_2)]$ and $0\leq\Theta \kappa_m M^2\leq1$. This nondimensionalization, where the $M^2$ factor appears explicitly, highlights the condition that, as the incompressible limit is approached following $M\rightarrow 0$, the background temperature (and hence density) becomes constant in each fluid region. Under these assumptions,  the background pressure is constant in time. Using the $T_{0m}$ variation with $x_1$ (Eq. \ref{eq:21}), Eq.(\ref{eq:17}) becomes
\begin{equation}\label{eq:25}
  \dD{p_{0m}}{x_1} = -\frac{\alpha_m M^2}{\Theta \kappa_m M^2 x_1 +1} p_{0m}\:. 
\end{equation}

\noindent
The solution to this equation is
\begin{equation}\label{eq:26}
  p_{0m} = p_{\infty}\left(\Theta \kappa_m M^2 x_1+1 \right) ^{-{\frac {\alpha_m}{\Theta \kappa_m}}}\:,
\end{equation} 
This solution is normalized so that the nondimensional pressure at the interface is $p_{\infty}$.
Using the equation of state, the density is obtained as
\begin{equation}\label{eq:27}
 \rho_{0m} = \alpha_m \left( \Theta \kappa_m M^2x_1+1
  \right) ^{-{\frac {\alpha_m}{\Theta \kappa_m}}-1} \:.
\end{equation}
The kinematic viscosity is then
\begin{equation}\label{eq:28}
\frac{\mu_{0m}}{\rho_{0m}} =\frac{\mu_{0m}}{\alpha_m}\left( \Theta \kappa_m M^2x_1+1  \right) ^{{\frac {\alpha_m}{\Theta \kappa_m}}+1} \:.
\end{equation}
In the subsequent analysis, a power law dependence of the dynamic viscosity with temperature, $\mu_{0m}=\mu_{0m,\infty} (\Theta M^2x_1+1)^{\xi}$, will be assumed, since this may be important in the practical applications considered,  as the temperature can have large variations across the RTI layer. The dimensionless isothermal sound speed can be written as
$c_m^2 = \frac{p_{0m}}{\rho_{0m}}=\frac{1}{\alpha_m} (\Theta \kappa_m M^2 x_1 + 1)$.
The equations reduce to those derived in Ref.~\cite{Livescu2004} when $\Theta\rightarrow 0$.

\subsection{First-order equations}
\label{sec:first}

The interface between the fluids is perturbed with an $x_2$ and
$x_3$ dependent perturbation.  The location of the
interface can be described using the function $x_s(x_2,x_3,t)$, with $\partial x_s
/ \partial t = u_1$. It is further assumed that the first order heat conduction term is small (large
$Pr\sqrt{Gr}$). Then the first-order linearized equations become
\begin{equation}\label{eq:33}
  \pD{\rho}{t} + \rho_0 \Delta + u_1 D \rho_0 = 0\:,
\end{equation}
\begin{subequations}
\begin{equation}\label{eq:34a}
  {\rho_0} \pD{u_1}{t} = -\frac{1}{M^2}Dp - \rho + 
\pD{}{x_j}\left[\frac{\mu_0}{\sqrt{Gr}}\left(\pD{u_1}{x_j} + 
Du_j\right)\right] - \frac{2}{3} D\left(\frac{\mu_0}{\sqrt{Gr}}\Delta\right)\:,
\end{equation}
 \begin{equation}\label{eq:34b}
{\rho_0} \pD{u_2}{t} = -\frac{1}{M^2}\pD{p}{x_2} + \pD{}{x_j}\left[\frac{\mu_0}{\sqrt{Gr}}\left(\pD{u_2}{x_j} + 
\pD{u_j}{x_2}\right)\right] - \frac{2}{3}\pD{}{x_2}\left(\frac{\mu_0}{\sqrt{Gr}}\Delta\right)\:,
\end{equation}
 \begin{equation}\label{eq:34c}
{\rho_0} \pD{u_3}{t} = -\frac{1}{M^2}\pD{p}{x_3}+ \pD{}{x_j}\left[\frac{\mu_0}{\sqrt{Gr}}\left(\pD{u_3}{x_j} + 
\pD{u_j}{x_3}\right)\right] - \frac{2}{3}\pD{}{x_3}\left(\frac{\mu_0}{\sqrt{Gr}}\Delta\right)\:,
\end{equation}
\end{subequations}
\begin{equation}\label{eq:35}
\pD{p}{t} = -\gamma p_0 \Delta- u_1 Dp_0\:.   
\end{equation}
In these equations $\Delta = \partial u_k/\partial x_k$, and $D =
\partial / \partial x_1$.
Following a normal mode analysis, solutions to these equations are sought with the $x_2$, $x_3$ and time dependencies of the form  $\exp(i(k_2x_2+k_3 x_3) + nt)$,
where $k_2 = \hat{k}_2\hat{L}$, $k_3 = \hat{k}_2\hat{L}$. $\hat{k}$ is defined as $\hat{k} = \sqrt{\hat{k}_2^2+\hat{k}_3^2}$, so that $k^2 = k_2^2+k_3^2$.
The growth rate, $\hat{n}$, is nondimensionalized as $n = \hat{n}\sqrt{\hat{L}/\hat{g}}$.

\section{Inviscid-inviscid interface}
\label{sec:3}

In absence of viscosity, $\mu = 0$; however, the background temperature gradient can still be present. In this case, if heat conduction is further assumed to be zero, the background temperature variation simply becomes $T=\Theta M^2 x +1$. The same relation is obtained for equal thermal conduction coefficients for the two fluids. After transforming equations (\ref{eq:33})-(\ref{eq:35}) into Fourier space, the equations for the amplitudes of the Fourier modes become (where the same notation was used for the real space variables and their Fourier amplitudes):

\begin{equation}\label{eq:39}
  n\rho = - \rho_0 \Delta - u_1 D \rho_0\:,
\end{equation}
\begin{equation}\label{eq:40}
  n \rho_0 u_1 = -\frac{1}{M^2}Dp -\rho\:,\ \  \rho_0 n u_2 = -\frac{1}{M^2}ik_2p\:, \  \ 
 \rho_0 n u_3 = -\frac{1}{M^2}ik_3 p\:,
\end{equation}
\begin{equation}\label{eq:41}
np = - \gamma p_0 \Delta + M^2 u_1 \rho_0\:.   
\end{equation}

After eliminating $p$, $\Delta$, $u_2$, and $u_3$ from these equations, an equation for $u_1$ is obtained as
\begin{equation}\label{eq:42}
\rho_0 u_1 = - u_1 D\left[\frac{\rho_0}{ n^2 + \gamma c^2k^2/M^2}\right]
+ D\left[\frac{\gamma c^2 \rho_0}{M^2(n^2+\gamma
    c^2k^2/M^2)}Du_1\right]+  \frac{u_1}{n^2}D\rho_0 + \frac{\rho_0
  u_1}{n^2 (n^2 + \gamma   c^2k^2/M^2)} \:.
\end{equation}
Eq.(\ref{eq:42}) gives the jump condition at the 
interface between the two fluids by integrating over an infinitesimal
element which includes the interface
\begin{equation}\label{eq:43}
- u_s  \delta\left[\frac{\rho_0}{ n^2 + \gamma c^2k^2/M^2}\right]
+ \delta\left[\frac{\gamma c^2 \rho_0}{M^2(n^2+\gamma
    c^2k^2/M^2)}Du_1 \right] + \frac{
  u_s}{n^2}\delta\rho_0 = 0\:,
\end{equation}
where the subscript $s$ denotes a quantity evaluated at the interface, whose location is given by the equation $x_s = x_1$, and $\delta f = f_+ - f_-$ with $f_+=f(x_s+0),$
$f_-=f(x_s-0)$.
After further simplifications, the equation for $u_1$ in each fluid (\ref{eq:42}) becomes
\begin{multline}\label{eq:44}
D^2u_1 - Du_1\left(\frac{ M^2}{c^2} + \frac{\gamma \Theta k^2}{\alpha
    (n^2+\gamma c^2k^2/M^2)}\right)
\\-u_1\left(k^2 + \frac{M^2 n^2}{\gamma c^2} +
  \frac{(\gamma-1)M^2k^2}{\gamma c^2 n^2} +
  \frac{\gamma \Theta k^4}{\alpha  n^2 (n^2+\gamma c^2k^2/M^2)}\right) =0
\:.
\end{multline}
Note that the coefficients in Eq.(\ref{eq:44}) are functions of
$x_1$ because the sound speed is a linear function of $x_1$. Eq. (\ref{eq:44}) does not admit an analytical  solution in a general case and is solved numerically using the following boundary conditions: $u_i=0$ at $x_1=\pm1$, continuity of $u_1$ at the interface, $\delta u_1=0$, and the jump condition Eq. (\ref{eq:43}). When $\Theta=0$, Eq.(\ref{eq:44}) becomes identical to the differential
equation derived in Ref.~\cite{Livescu2004}.  

In the limit of large $\Theta$, the coefficients in Eq.(\ref{eq:44}) are dominated by
temperature gradient effects and the equation reduces to
\begin{equation}\label{eq:47}
D^2u_1 - \frac{\gamma \Theta k^2}{\alpha ( n^2 + \gamma c^2 k^2/ M^2)} Du_1 - \bigg(k^2+ \frac{\gamma \Theta k^4}{\alpha n^2 ( n^2 + \gamma c^2 k^2/ M^2)} \bigg) u_1 = 0.
\end{equation}
The solution to Eq.(\ref{eq:47}) is
\begin{multline}\label{eq:48}
u_1 = e^{-k\left( x_1+\frac{\alpha n^2}{\gamma \Theta k^2}+\frac{1}{\Theta M^2} \right)}\left(\frac{x_1}{n^2}+\frac{\alpha}{\gamma \Theta k^2}+\frac{1}{\Theta M^2 n^2}\right)^2  \bigg[ C_1 U\left(\frac{3+k/n^2}{2}, 3, 2k\left( x_1+\frac{\alpha n^2}{\gamma\ \Theta k^2}+\frac{1}{\Theta M^2}\right) \right) \\ + C_2 L\left( - \frac{3+k/n^2}{2}, 2, 2k\left( x_1+\frac{\alpha n^2}{\gamma\ \Theta k^2}+\frac{1}{\Theta M^2}\right) \right) \bigg],
\end{multline}
 where: $U$ is the confluent hypergeometric Kummer's function of the second kind and $L$ is the associated Laguerre's polynomial. The coefficients $C_1$, $C_2$ are determined to a multiplying constant from the conditions that $u_1$ vanishes at the rigid boundaries located at $x_1=\pm1$ and that it is continuous over the interface. After replacing $u_1$ in the jump condition, a dispersion equation for the  growth rate can be obtained (not shown in the paper because of its cumbersomeness).
 
In the incompressible limit ($M^2=0$), the dispersion relation simplifies to an explicit formula for the growth rate, $n^2/k=At \tanh{k}$, which corresponds to the finite domain growth rate equation from Ref.~\cite{Livescu2004}. In this case, the normalized growth rate becomes zero in the limit of small domain size with respect to the perturbation wavelength ($k\rightarrow 0$) and approaches the infinite domain formula ($n^2/k=At$) in the limit of large domain size with respect to the perturbation wavelength 
($k\rightarrow \infty$).

\section{Viscous-viscous interface}
\label{sec:4}

For the viscous case, neglecting viscosity fluctuations in $x_2$ and $x_3$ so that viscosity varies in $x_1$ direction only, the equations (\ref{eq:33}) -- (\ref{eq:35}) become
\begin{equation}\label{eq:49}
  n\rho = - \rho_0 \Delta - u_1 D \rho_0\:,
\end{equation}
\begin{subequations}
\begin{equation}\label{eq:50a}
  n \rho_0 u_1 = -\frac{Dp}{M^2} - \rho + \frac{2 D \mu_0}{\sqrt{Gr}}  \left( Du_1 - \frac{\Delta}{3} \right) +\frac{\mu_0}{\sqrt{Gr}} \left( D^2u_1 - k^2u_1+\frac{D\Delta}{3} \right) \:,
\end{equation}
 \begin{equation}\label{eq:50b}
n \rho_0 u_2 = -\frac{ik_2p}{M^2} + \frac{D \mu_0}{\sqrt{Gr}}  \left( Du_2 - ik_2u_1 \right) +\frac{\mu_0}{\sqrt{Gr}} \left( D^2u_2 - k^2u_2+\frac{ik_2\Delta}{3} \right) \:,
\end{equation}
 \begin{equation}\label{eq:50c}
n \rho_0 u_3 = -\frac{ik_3p}{M^2} + \frac{D \mu_0}{\sqrt{Gr}}  \left( Du_3 - ik_3u_1 \right) +\frac{\mu_0}{\sqrt{Gr}} \left( D^2u_3 - k^2u_3+\frac{ik_3\Delta}{3} \right) \:,
\end{equation}
\end{subequations}
\begin{equation}\label{eq:51}
np = - \gamma p_0 \Delta + M^2 u_1 \rho_0\:.   
\end{equation}
Following a similar procedure as in the previous section, the equation for $u_1$ is obtained as a fourth order ordinary differential equation
\begin{equation}\label{eq:52}
A_4D^4u_1 + A_3D^3u_1 + A_2D^2u_1 + A_1Du_1 + A_0u_1 = 0 \:,
\end{equation}
where the coefficients $A_i$ are given in Appendix. The boundary conditions for Eq.(\ref{eq:52}) are: vanishing velocity at the rigid boundaries, $u_i=0$ and $\Delta - Du_1=0$ at $x_1=\pm1$, continuity of velocity and tangential stress at the interface, $\delta u_1=0$, $\delta (\Delta - Du_1)=0$ and  $\delta [\mu_0 (D\Delta - D^2u_1 - k^2u_1)]=0$ at $x_s=x_1$. The jump condition is
\begin{multline}\label{eq:53}
\delta \left[ \left( -\rho_0 +\frac{\mu_0}{n\sqrt{Gr}}D^2 \right) \left( \Delta - Du_1 \right) \right] 
+ \frac{k^2}{n\sqrt{Gr}}\delta \left( \mu_0Du_1 \right) + \frac{1}{n\sqrt{Gr}}\delta \left( D\mu_0 \right) (D\Delta - D^2u_1 - k^2u_1)_s\\
= - \frac{k^2}{n^2}\delta(\rho_0)u_{1,s} + \frac{2k^2}{n\sqrt{Gr}}\delta(\mu_0)(\Delta - Du_1)_s \:.
\end{multline}
The divergence of velocity, $\Delta$, is given by 
\begin{equation}\label{eq:54}
\beta_1\Delta = B_3D^3u_1 + B_2D^2u_1 + B_1Du_1 + B_0u_1 \:,
\end{equation}
with the coefficients $\beta$ and $B_i$ and expressions for $D \Delta$ and $D^2\Delta$ in terms of the derivatives of $u_1$ are provided in the Appendix. Since $u_1$ can be found only to a multiplying constant, the boundary conditions are supplemented with the specification of $u_1$ or one of its derivatives at one point inside of the domain. Then Eq.(\ref{eq:52}) with the boundary and jump conditions form a closed set of equations from which $u_1$ on each side of the interface and the growth rate, $n$, can be determined. Eq.(\ref{eq:52}) is numerically integrated on each side of the domain using a fourth order Runge-Kutta scheme. In order to determine $n$ and $u_1$ from the matching conditions at the interface, a multidimensional secant method (Broyden's method) is employed. \cite{Press1986} This numerical method, where the equations are integrated starting from one boundary to the next works very well at small to moderate $Gr$, but it can become unstable at large $Gr$ values. An approach which can capture the case when $Gr$ is large for one of the fluids is described in the next section.

\section{Viscous-inviscid interface}
\label{sec:7}

In some practical applications (including the two applications considered here), the viscosity ratio between the two fluids is large enough so that, for the range of wavenumbers around the most unstable mode corresponding to one of the fluids, the other fluid has negligible viscous effects. In this case, the first fluid needs to still be considered as viscous, while the second can be treated as inviscid. This allows the equations to simplify considerably compared to the fully viscous case and also the use of the numerical integration method described in the previous section. For ICF and solar corona examples, the large viscosity ratio between the two fluids is due to the very large temperature difference between the hot spot and DT ice during the ICF coasting stage and solar coronal plasma and prominence plumes, respectively. 
Consistent with these two examples, here we consider that the light fluid is viscous and the heavy is inviscid. Then the boundary conditions  are: $u_i=0$ at $x_1=\pm1$,  vanishing tangential velocity  at the rigid boundary only for the viscous side, i.e. $\Delta - Du_1=0$ at $x_1=-1$, continuity of $u_1$  at the interface, $\delta u_1=0$ at $x_1=x_s$,  and vanishing viscous tangential stress  at the interface due to slip condition, $\mu_0 (D\Delta - D^2u_1 - k^2u_1)=0$ at $x_s=x_1$. Unlike the viscous-viscous case, the tangential velocity is not continuous at the interface.  The jump condition becomes
\begin{multline}\label{eq:55}
\delta \left[ \left( -\rho_0 +\frac{\mu_0}{n\sqrt{Gr}}D^2 \right) \left( \Delta - Du_1 \right) \right] 
+ \frac{k^2}{n\sqrt{Gr}}\delta \left( \mu_0Du_1 \right) \\
= - \frac{k^2}{n^2}\delta(\rho_0)u_{1,s} + \frac{2k^2}{n\sqrt{Gr}}\delta(\mu_0(\Delta - Du_1)) \:,
\end{multline}
where $\mu_0=0$ on the inviscid part and $Gr$ corresponds to the viscous part of the interface.

In the incompressible limit ($p_\infty \rightarrow \infty$) and without temperature gradient ($\Theta=0,\ M^2=0$), 
a dispersion equation for the growth rate can be obtained. Eq.(\ref{eq:52}) for the viscous part in such case reduces to
\begin{equation}\label{eq:56}
D^4u_1 - (n\sqrt{Gr}+2k^2)D^2u_1 + (n\sqrt{Gr}+k^2)k^2u_1 = 0 
\end{equation}
and can be analytically solved
\begin{equation}\label{eq:57}
u_1 = C_1e^{qx_1} +  C_2e^{-qx_1} + C_3e^{kx_1} + C_4e^{-kx_1} \:,
\end{equation}
where $q=\sqrt{n\sqrt{Gr}+k^2}$.
Eq.(\ref{eq:44}) for the inviscid part becomes
\begin{equation}\label{eq:collected}
D^2u_1 - k^2u_1 = 0 
\end{equation}
and has the solution
\begin{equation}\label{eq:59}
u_1 = C_5e^{kx_1} +  C_6e^{-kx_1} \:,
\end{equation}
where $C_1$ through $C_6$ are constants of integration.
After applying the boundary and jump conditions to Eq.(\ref{eq:57}) and (\ref{eq:59}) to eliminate the constants of integration, a dispersion equation  can be obtained as
\begin{multline}\label{eq:60}
\left( qC_qS_k-kC_kS_q \right) \left[ 4k^2R_1S_k+ \left( q^2-k^2 \right) \left( R_5e^{-k}-R_6e^k \right) \right]
\\+(R_2-R_1)S_k \left[ 2k^2e^{-q} \left( kC_k+qS_k \right)-k \left( q^2+k^2 \right) \right]+2R_3S_k \left(q^2+k^2 \right) \left(kC_kS_q-qC_qS_k \right)
\\+(R_4-R_3)S_k \left[ e^{-k} \left( q^2+k^2 \right) \left( qC_q+kS_q \right) -2qk^2 \right] =0 \:,
\end{multline}
where the coefficients are defined as
$R_1,R_2=At/(2n^2)\mp 2\alpha_1q/(n\sqrt{Gr})$;
$R_3,R_4=At/(2n^2)\mp 2\alpha_1k/(n\sqrt{Gr})\mp \alpha_1/k$;
$R_5,R_6=At/(2n^2)\pm \alpha_2/k$;
$C_q,S_q=0.5(e^q\pm e^{-q})$;
$C_k,S_k=0.5(e^k\pm e^{-k})$, with upper and lower signs corresponding to the left and right coefficients, respectively.

For  a large domain compared to the wavelength of the perturbation ($k=\hat{k}\hat{L}\gg 1$), keeping only the dominant terms in Eq.(\ref{eq:60}) simplifies this equation to
\begin{equation}\label{eq:61}
\frac{4\alpha_1 k^2}{Gr} \left( -k\sqrt{n\sqrt{Gr}+k^2}+k^2+n\sqrt{Gr}  \right)-kAt+n^2=0 \:,
\end{equation}
where  $At=(\rho_{02}-\rho_{01})/(\rho_{02}+\rho_{01})$. For the case of infinitely small viscosity on the viscous side ($Gr \rightarrow \infty$), the first term in Eq.(\ref{eq:61}) becomes small compared to the other two terms and  the equation reduces to the classical inviscid-inviscid incompressible interface  for an infinite domain $n^2 / k = At$, or in the dimensional form $\hat{n}^2 / \hat{g}\hat{k} = At$. \cite{Chandrasekhar1981}

\section{Discussion}
\label{sec:8}

\begin{figure}[ht]
\includegraphics[width=5in]{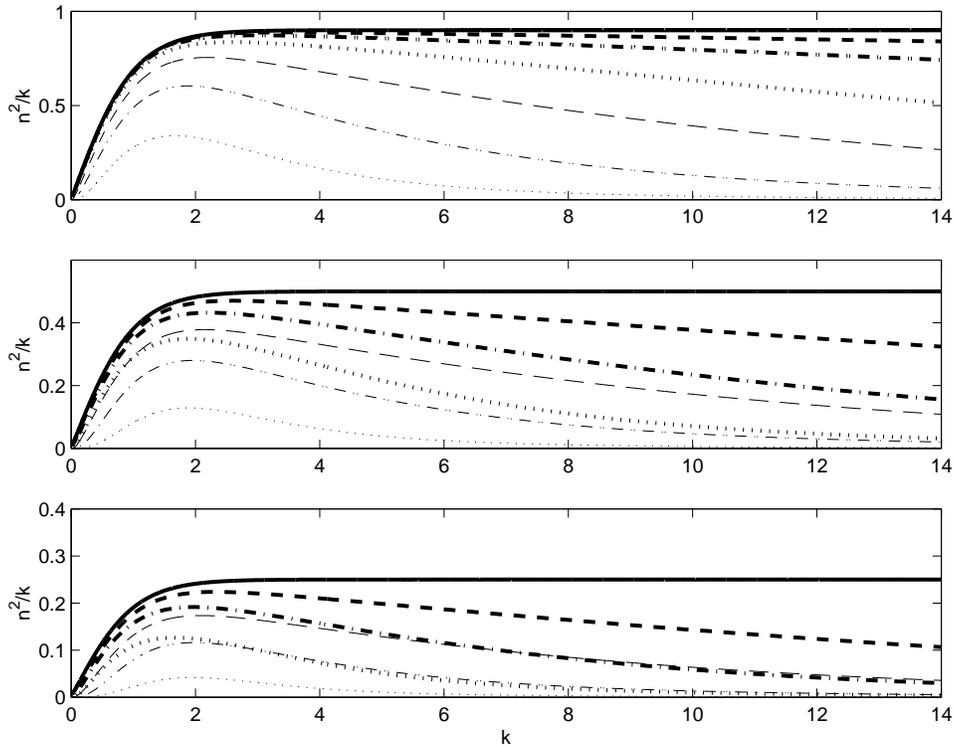}
\caption{Growth rate normalized  by  wave number $n^2/k=\hat{n}^2/\hat{k}\hat{g}$ for $M^2=0$, $\Theta=0$. Plots from top to bottom: $At=0.9, \: 0.5, \: 0.25$.
Thick solid line: inviscid-inviscid, 
thin lines: viscous-viscous,
thick non-solid lines: viscous-inviscid.
Dashed line: $Gr=10,000$, dashed-dot line: $Gr=1000$, dot line: $Gr=100$.}
\label{fig:visc5}
\end{figure}

The comparison of all three interface types for different $At$ values in the simplest incompressible case without temperature gradient ($M^2=0$, $\Theta=0$) is shown in  Fig.~\ref{fig:visc5}. As $Gr$ increases, the normalized growth rates obtained for the viscous cases approach the limiting inviscid case results for some low $k$ range which increases with $At$. Viscous cases have a most unstable mode close to $k\approx2$ (the wavenumber location decreases with $At$), which is relatively insensitive to $Gr$. While for all viscous and viscous-inviscid cases the normalized growth rate goes to zero as $k\rightarrow \infty$, it is interesting to mention the growth behavior around the most unstable mode. When one side is inviscid, the results are very close to the fully inviscid case at high $At$, with larger differences at low $At$. Compared to the viscous-viscous case, there are noticeable differences for the viscous-inviscid case results at all $At$ values. 

\begin{figure}[ht]
\includegraphics[width=5in]{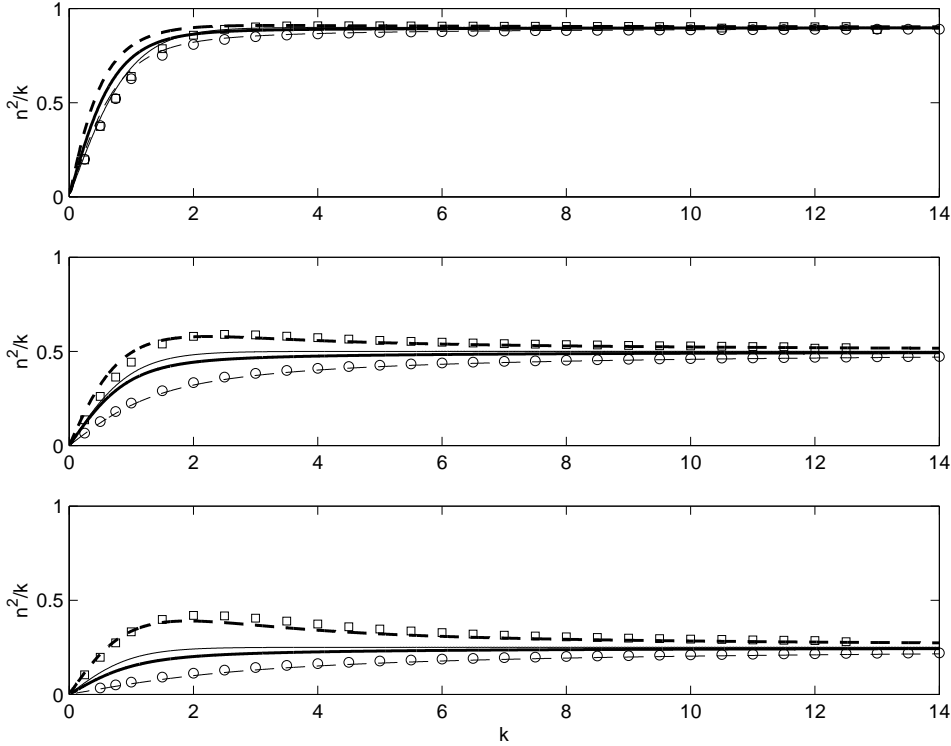}
\caption{Growth rate for the inviscid-inviscid case. Plots from top to bottom: $At=0.9, \: 0.5, \: 0.25$.
Dashed thick line: $\Theta =-0.9$, $M^2=1$, 
solid thin line:  $\Theta=0$, $M^2=0$,
solid thick line:  $\Theta=0$, $M^2=1$,
dashed thin line: $\Theta=0.9$, $M^2=1$, 
squares:   $\Theta =-0.9$, $M^2=1$ from Eq.(\ref{eq:47}), 
circles:   $\Theta =0.9$, $M^2=1$ from Eq.(\ref{eq:47}).}
\label{fig:inviscid1}
\end{figure}

\begin{figure}[ht]
\includegraphics[width=5in]{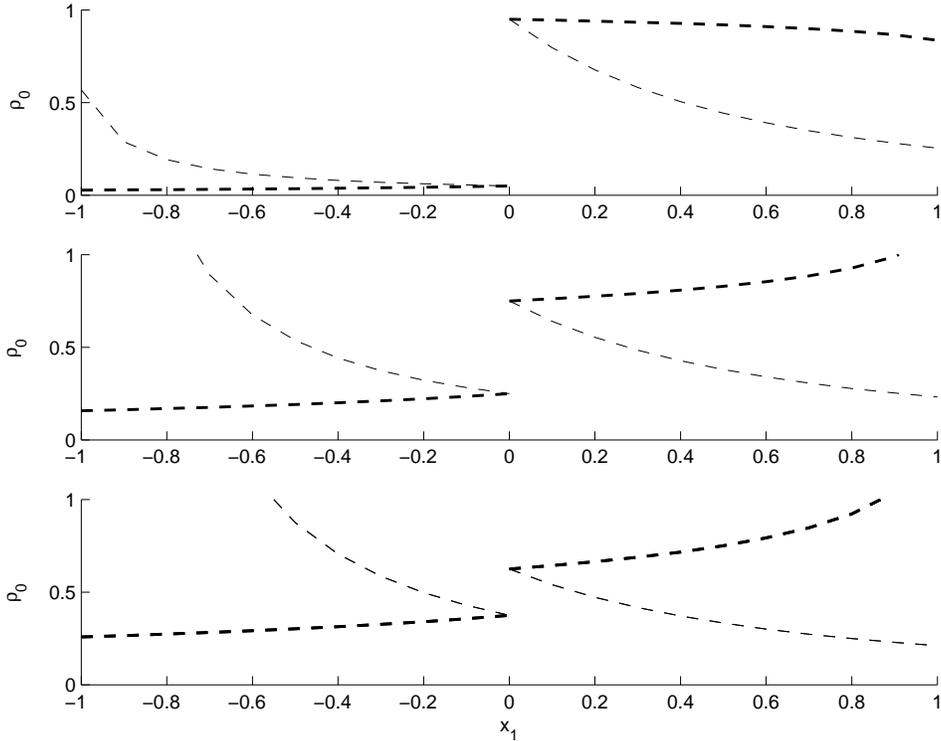}
\caption{Background density profiles, $\rho_{0} = \alpha \left( \Theta M^2x_1+1
  \right) ^{-{\frac {\alpha}{\Theta}}-1}$. Plots from top to bottom: $At=0.9, \: 0.5, \: 0.25$.
Dashed thick lines: $\Theta=-0.9$, $M^2=1$, 
dashed thin lines: $\Theta=0.9$, $M^2=1$.}
\label{fig:inviscid3}
\end{figure}

The effect of $\Theta\neq 0$ on the normalized growth rate for the inviscid-inviscid case is shown in Fig.~\ref{fig:inviscid1}. For the compressible cases, we chose $M^2=1$ as representative. The role of $M^2$ on the instability growth was discussed in Ref.~\cite{Livescu2004} Negative background temperature gradients ($\Theta<0$), corresponding to hotter light fluid, yield growth rates larger than those obtained for $\Theta=0$. Thus, the effect should be destabilizing for the two applications considered in this paper. Both the negative and positive temperature gradient effects are more pronounced for smaller $At$ and smaller $k$ ($k<6$). This is consistent with the variation of the background stratification subsequent to the variation of $\Theta$ (Fig. \ref{fig:inviscid3}). Thus, the integral Atwood numbers, $At_I$, presented in Table~\ref{TableParameters}, show larger differences between positive and negative $\Theta$ cases at small nominal $At$. Here, $At_I$ values are calculated using the background density integrals over the heavy and light fluid regions, respectively. 

While the growth rates obtained for $\Theta<0$ are larger than those obtained for the $\Theta=0$ compressible case for all $k$ values, they also become larger than those obtained for the\begin{table*}[ht]
\begin{tabular}{|  c  ||  c  |  c  |  c  |  c |}  
\hline
    Case & $At_{I-}$ & $At_{I+}$ & $\delta At_{I\pm}$ & $\delta n^2/k_{\pm}$ \\
\hline
$At=0.9$ & 0.92 & 0.56 & 0.36 & 0.08 \\
$At=0.5$ & 0.63 & --0.39 & 1 & 0.26 \\
$At=0.25$ & 0.45 & --0.65 & 1.1 & 0.3 \\
\hline
\end{tabular}
\caption{Integral Atwood numbers, $At_{I\pm}=(\rho_{2,I\pm}-\rho_{1,I\pm})/(\rho_{2,I\pm}+\rho_{1,I\pm})$, and their differences, $\delta At_{I\pm}=At_{I-}-At_{I+}$. The integral densities, $\rho_{m,I}$, are calculated by integrating the background density profiles over the two fluid regions. The differences between the growth rates, $\delta n^2/k_{\pm}\equiv n^2/k_{-}-n^2/k_{+}$,  are calculated from Fig.~\ref{fig:inviscid1} at $k\approx2$. $\pm$  refer to $\Theta=0.9$ and $\Theta=-0.9$ cases, respectively. $M^2=1$.} 
\label{TableParameters}
\end{table*} incompressible constant density case in an infinite domain ($n^2/k=At$) for $k$ sufficiently large. Again, this is consistent with the background density variation and the fact that $\Theta<0$ growth rates are larger than $\Theta=0$ growth rates and the latter should approach the incompressible constant density infinite domain results as $k\rightarrow \infty$.\cite{Livescu2004} Since the normalized growth rate starts from small values and approaches the asymptotic value $n^2/k=At$ from above as $k$ increases, there is a maximum normalized growth rate which, interestingly, occurs around $k\approx 2$, similar to the most unstable mode obtained for the viscous cases. Again, this effect is more pronounced at small $At$ and becomes negligible as $At$ approaches $1$. 

The results for the growth rate obtained analytically from Eq.\ref{eq:47}  in the large temperature gradient limit are also presented in Fig.~\ref{fig:inviscid1}. The analytical formula follows the numerical results for a finite temperature gradient and approximates these results well for small $At$ and/or large $k$ values (large domain size compared to the perturbation wavelength).

Figure~\ref{fig:inviscid1} also shows that the compressible growth rate with $\Theta=0$ can become larger than the corresponding incompressible growth rate at $At=0.9$ and $k \lessapprox 2$. This overshoot occurs in a different parameter range than that studied in Ref.~\cite{Livescu2004} Nevertheless, unlike the overshoot studied in Ref.~\cite{Livescu2004}, in this case, since the normalized growth rates for compressible and incompressible $\Theta=0$ cases increase monotonically with $k$, the infinite domain incompressible constant density growth rate still represents the upper bound.\begin{figure}[ht]
\includegraphics[width=5in]{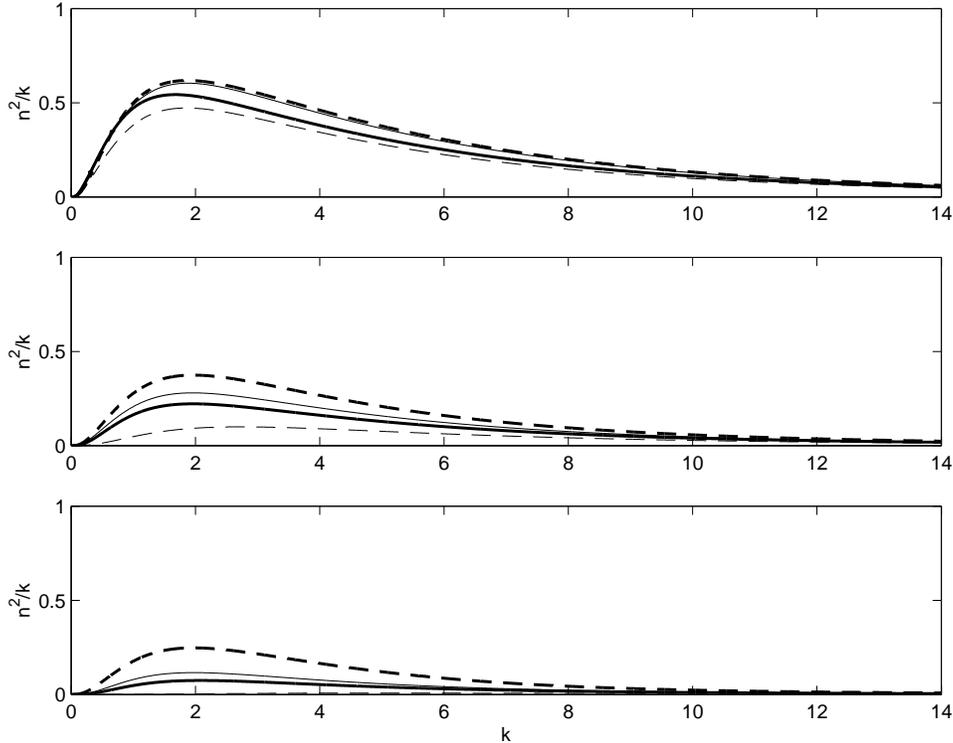}
\caption{Growth rate  for viscous-viscous case computed at $Gr=1000$. Plots from top to bottom: $At=0.9, \: 0.5, \: 0.25$.
Dashed thick line:  $\Theta=-0.9$, $M^2=1$,
solid thin line:   $\Theta=0$, $M^2=0$,
solid thick line:  $\Theta=0$, $M^2=1$,
dashed thin line:  $\Theta=0.9$, $M^2=1$. }
\label{fig:visc2}
\end{figure} 

\begin{figure}[ht]
\includegraphics[width=5in]{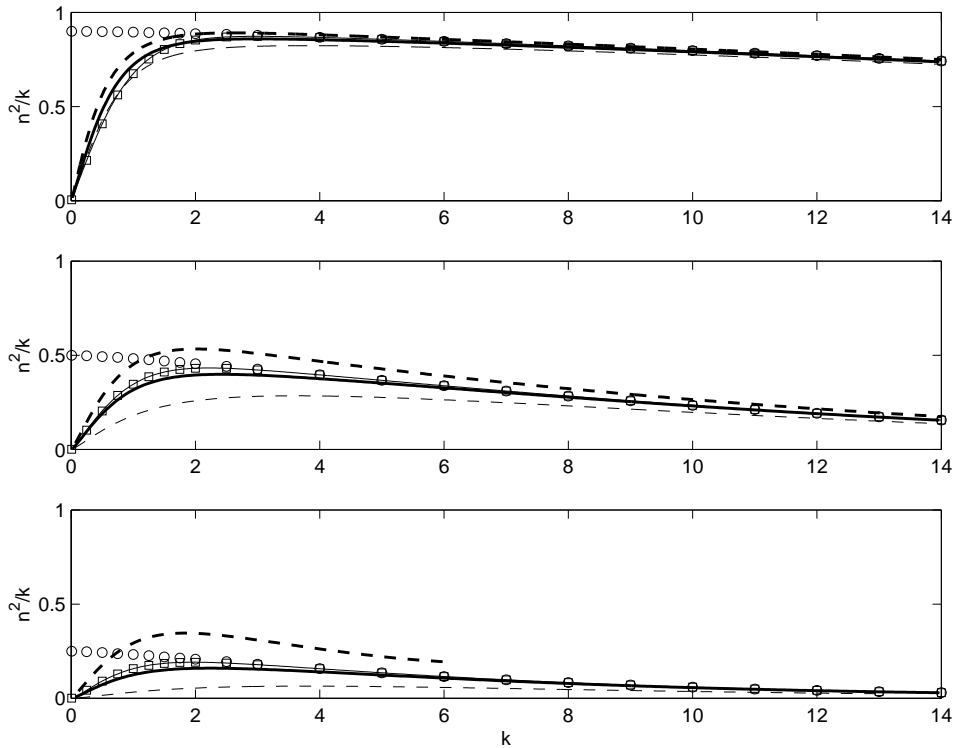}
\caption{Growth rate  for viscous-inviscid case computed at $Gr=1000$. Plots from top to bottom: $At=0.9, \: 0.5, \: 0.25$.
Dashed thick line: $\Theta=-0.9$, $M^2=1$,
solid thin line:   $\Theta=0$, $M^2=0$,
solid thick line:   $\Theta=0$, $M^2=0$,
dashed thin line:  $\Theta =0.9$, $M^2=1$.
Squares: from Eq.(\ref{eq:60}) for the incompressible, uniform background temperature case,
circles: from Eq.(\ref{eq:61}) for the large domain size compared to the perturbation wavelength case. }
\label{fig:viscinvis1}
\end{figure}

The effects of the background temperature gradient on the normalized growth rate for viscous-viscous and viscous-inviscid (lower part / light fluid -- viscous, upper part / heavy fluid-- inviscid) cases are shown in Fig.~\ref{fig:visc2} and Fig.~\ref{fig:viscinvis1}, respectively. Viscosity is important at all scales and dominates at large $k$, when the normalized growth rate asymptotes to zero. Similar to the inviscid-inviscid interface, a $\Theta <0$, corresponding to hotter light fluid, has a destabilizing effect, while $\Theta>0$ has a stabilizing effect. The destabilizing effect of $\Theta<0$ is more pronounced at low $At$ and becomes smaller as $At \rightarrow 1$.  
The peak of $n^2/k$ (most unstable mode) with respect to its location corresponding to the $\Theta M^2=0$ case shifts to larger $k$ values for $\Theta>0$ and to lower $k$ values for $\Theta<0$. This effect becomes more pronounced for smaller $At$ values. The results are qualitatively similar for the viscous-inviscid case, though these are closer to the $\Theta=0$ results at high $At$ values. For the viscous-inviscid case, Fig.~\ref{fig:viscinvis1} also shows the results obtained from the dispersion relations Eq.(\ref{eq:60}) and Eq.(\ref{eq:61}) derived  for incompressible, constant temperature case. The dispersion relation (\ref{eq:60}) gives identical results with those calculated numerically by integrating the governing ordinary differential equations. The simplified formula (\ref{eq:61}) does not produce the correct $k\rightarrow 0$ limit, as expected, but becomes a good approximation for $k\gtrapprox 2$.
The growth rate for inviscid-viscous case (lower part / light fluid-- inviscid, upper part / heavy fluid -- viscous) is qualitatively similar but slightly smaller in magnitude than for viscous-inviscid interface (not shown here). 

\section{Application to ICF and solar corona}
\label{sec:ICF}

In this section, the viscous-viscous and viscous-inviscid formulas are applied to two practical situations: the coasting phase in ICF and solar corona plumes. While these applications contain many other complicating plasma physics effects, we demonstrate what a normal mode analysis using two immiscible fluids predicts for the range of parameters associated with such applications. Results from normal mode analysis with immiscible fluids have been routinely used for these applications to qualitatively predict the importance of RTI and, as far as we know, this is the first time when such an analysis is used with viscous (or viscous-inviscid), compressible fluids with a background temperature gradient. We further assume that  the density variation across the domain is concentrated at the interface such that the nominal $At$ numbers are matched. 

In ICF, ignition is triggered by a hot spot at the center of an imploded fuel shell. \cite{Lindl1998, Atzeni2004} The hot spot formation requires implosion symmetry, which is hindered, in particular, by the development of RTI. The focus of the present study is the instability that forms during the coasting or deceleration phase, before stagnation, when the dense fuel shell is decelerated by the pressure exerted by the hot and less dense inner plasma. Curvature effects are neglected in this analysis; this implies that the results are not applicable at small wavenumbers, i.e. $\hat{k} \lesssim 1/R$ (R -- radius of ICF shell).

\begin{figure}[ht]
\includegraphics[width=5in]{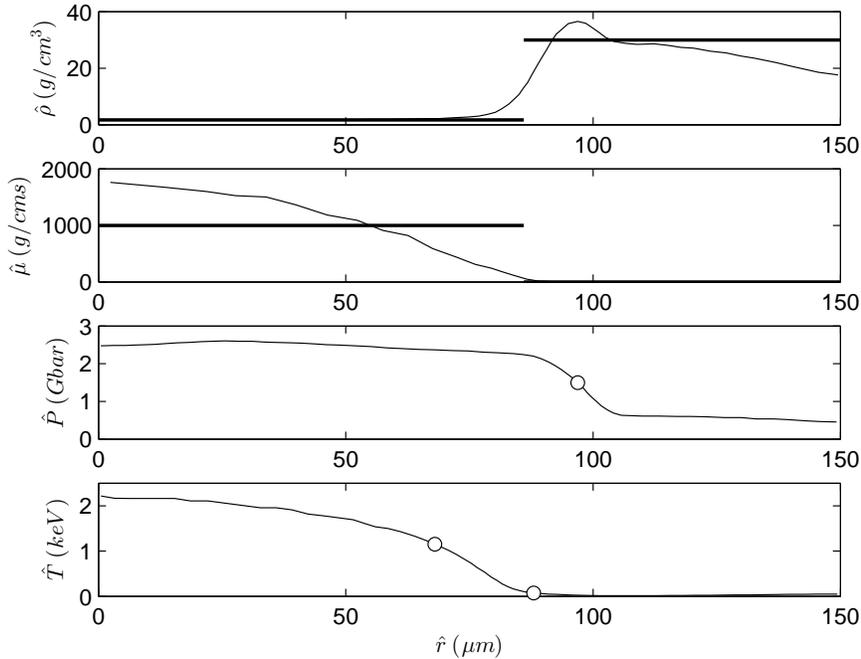}
\caption{Profiles from Weber et al. (Fig. 3). \cite{Weber2014}
Thick lines in the density and viscosity profiles: jumps applied in the present simulations.
Circle in the pressure profile: pressure $\hat{P}_\infty=1.5Gbar$  applied in the present simulations.
Circles in the temperature profile are the  temperatures on both sides $\hat{T}_1=1.15keV$ and $\hat{T}_2=0.075keV$ applied in the present simulation and corresponding to  
$\Theta =(\hat{T}_2- \hat{T}_1)/(\hat{T}_2+ \hat{T}_1)=-0.88$ with $\hat{T}_\infty = (\hat{T}_2+ \hat{T}_1)/2= 0.6keV$.}
\label{fig:icf1}
\end{figure}

The input parameters for  the numerical calculations are taken from Weber et al. \cite{Weber2014} Fig.~\ref{fig:icf1} shows the radial profiles, averaged over $4\pi$, corresponding to the beginning of the coasting phase. In their computational study, Weber et al. \cite{Weber2014} considered the importance of plasma viscosity on the development of turbulent mixing during the coasting phase. In another computational study using ILES (implicit Large Eddy Simulation), Haines et al. \cite{Haines2014} investigated the effects of plasma viscosity and diffusion on the turbulent instability growth under the ICF conditions. The main conclusion reached in these references is that the small scale turbulent motions resulting from RTI are damped by the increased viscosity in the hot spot, while only the large scale perturbations, possibly carrying over the imprint of laser non-uniformities, survive. The normal mode analysis described here does not account for mass diffusion, but it is still interesting to see what it predicts for the range of wavenumbers likely to survive the damping due to increased viscosity of the hot fluid. 

\begin{figure}[ht]
\includegraphics[width=5in]{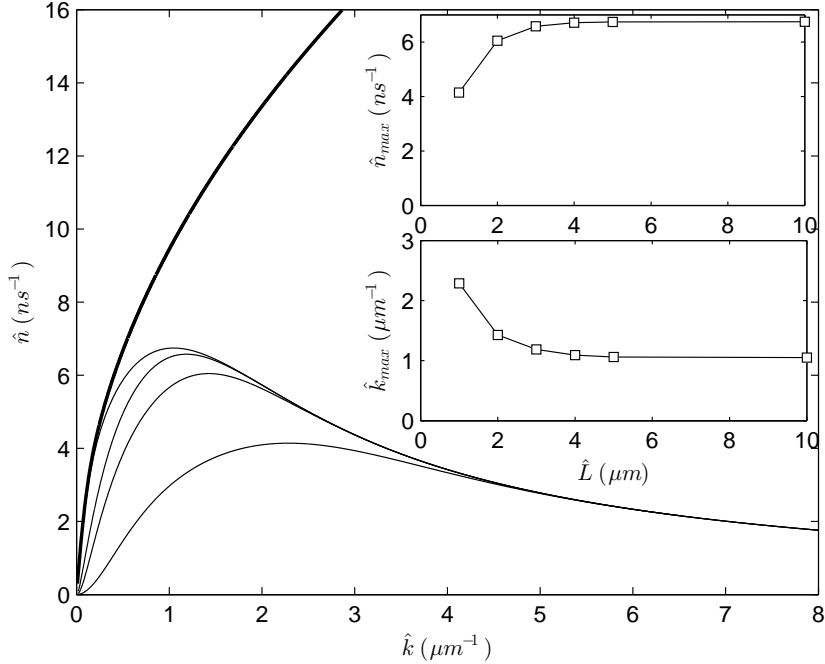}
\caption{The dimensional growth rates  calculated based on the parameters from Fig.~\ref{fig:icf1}  in the viscous-inviscid interface limit: $At=0.9$,  $\Theta=0$, $\hat{g}=10^{14} m/s^2$, $\gamma_1=\gamma_2=5/3$. Solid thin lines from top to bottom: $\hat{L}=10\mu m$, $3\mu m$, $2\mu m$, $1\mu m$, for $\hat{\mu}_1=1000 g/cm s$, $\hat{\mu}_2=0 g/cm s$,
thick solid line: inviscid-inviscid interface at $\hat{L}=10\mu m$. $\hat{L}=10\mu m$ corresponds to $M^2\approx0.2$, $Gr\approx10$.
Inset: peak value of the dimensional growth rate  (upper) and its location (lower) as functions of the domain size.}
\label{fig:icf2}
\end{figure}

One degree of uncertainty related to the set of parameters described above is the vertical extent of the domain. As explained in section \ref{sec:4}, the numerical integration method used here becomes unstable at large $Gr$ values and increasing the domain size has the effect of increasing the overall $Gr$. On the other hand, small domain sizes are affected by finite size effects and yield lower growth rates. Thus, first the influence of $\hat{L}$ is examined. Fig.~\ref{fig:icf2}  shows the growth rate in dimensional form in the approximation of viscous-inviscid interface and uniform background temperature for different $\hat{L}$ values. The inviscid-inviscid case is also presented for comparison. The growth rate increases significantly as $\hat{L}$ increases from $1\mu m$ to $10\mu m$; however, it appears to converge as $\hat{L}$ approaches $~10\mu m$. The inset to Fig.~\ref{fig:icf2}  shows the peak value and the peak location of the growth rate as a function of $\hat{L}$. For $\hat{L} > 5\mu m$ both the peak and its location almost do not depend on the domain size, which means that calculations made with the domain size $\hat{L} = 10\mu m$ provide converged results. It can also be concluded from the comparison of the viscous-inviscid and inviscid-inviscid interfaces that  viscosity has a strong damping effect on RTI during the coasting phase.

\begin{figure}[ht]
\includegraphics[width=5in]{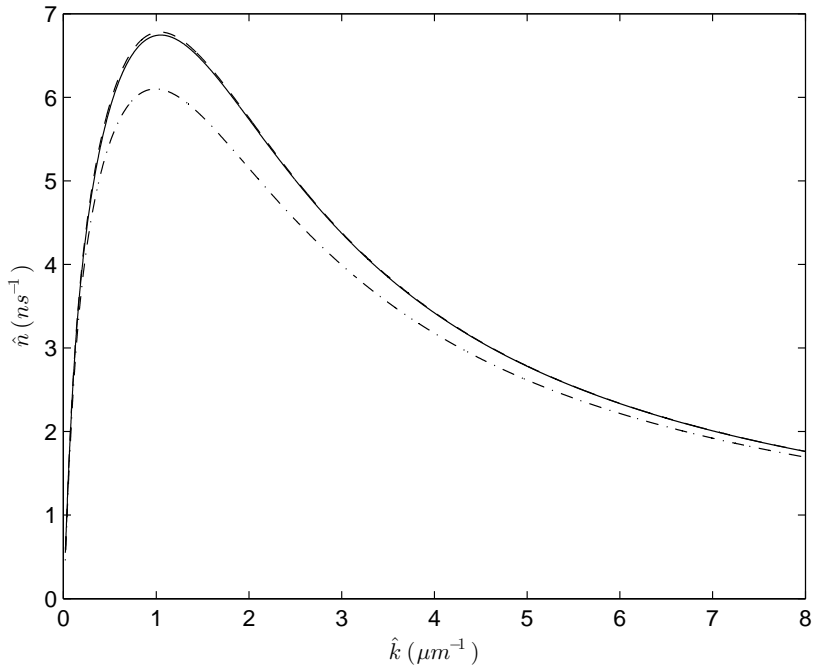}
\caption{Dimensional growth rates calculated based on the parameters from Fig.~\ref{fig:icf1} in the viscous-inviscid interface limit: $\hat{L}=10\mu m$ and $\hat{\mu}_1=1000$ g/cm s, $\hat{\mu}_2=0 g/cm s$.  
Solid line: constant temperature $\Theta=0$, $M^2\approx0.2$,
dashed line: $\Theta=-0.88$, $M^2\approx0.2$, $\xi=0$,
dashed-dot line: $\Theta=-0.88$, $M^2\approx0.2$, $\xi=2.5$.}
\label{fig:icf4}
\end{figure}

Fig.~\ref{fig:icf4} presents the effect of $\Theta$ on the growth rate.  The  temperature gradient is estimated from  Fig.~\ref{fig:icf1} and corresponds to non-dimensional value $\Theta=-0.88$. $\Theta \neq 0$ adds  only about $0.5 \%$ to the peak growth rate  in the case of  temperature independent viscosity. Using temperature dependent viscosity, $\hat{\mu}  = \hat{\mu}_{0,\infty}(\hat{T}/\hat{T}_\infty)^{2.5}$,  where the exponent follows the formulas derived in Braginksii~\cite{braginskii1965}, reduces the peak more noticeably by about $10\%$. The wavenumbers least affected by viscosity and showing significant growth are around the most unstable mode and lie in the range $\hat{k}\approx0.1\mu m^{-1}$ to $\hat{k}\approx8\mu m^{-1}$.

The results can be qualitatively compared with previous results related to the ICF coasting. \cite{Betti1998, Betti2001,Atzeni2004a,Bychkov2015}. These studies include the effect of ablation, however they do not account for the presence of viscosity. In general, the growth rate calculated for viscous-inviscid compressible interface at $\hat{L} = 10\mu m$ is less than the values obtained in these studies, pointing to the importance of including the physical transport in the multi-dimensional calculations.

In ICF, RTI grows at micron scale. It would be interesting to compare the results to large scale applications. As an example, the solar corona is considered here where the instabilities can develop on hundreds to thousands kilometers scale. RT-type instabilities are formed at the interface of the quiescent low density coronal plasma and the prominence plumes of denser plasma from the chromosphere, providing $At=0.6 \div 0.7$ and $\Theta \approx -0.9$. \cite{Berger2010, Ballai2015, Delcroix1968} Because the prominence plasma is strongly magnetized, there is a dumping effect of magnetic pressure on the instability growth, that, in addition to other complex plasma properties, is not taken into account in the present estimates.\begin{figure}[ht]
\includegraphics[width=5in]{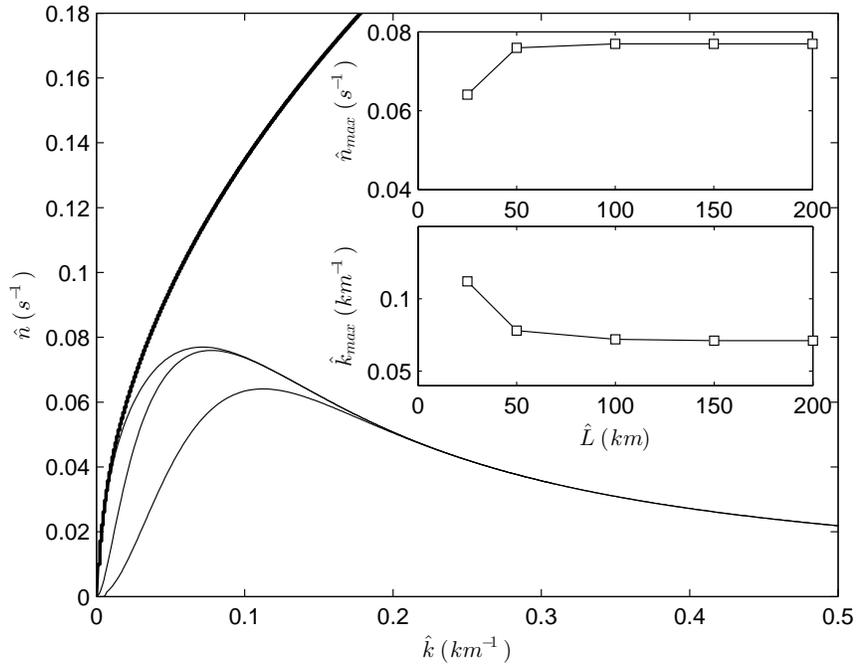}
\caption{The dimensional growth rates for the solar corona calculated in the  viscous-inviscid interface limit: $At=0.67$,  $\Theta=0$, $\hat{g}=0.274 km/s^2$, $\gamma_1=\gamma_2=5/3$. 
Solid thin lines from top to bottom: $\hat{L}=200 km$, $50 km$, $25 km$ for $\hat{\mu}_1=10 kg/km s$, $\hat{\mu}_2=0 kg/km s$,
thick solid line: inviscid-inviscid interface at $\hat{L}=200 km$. $\hat{L}=200km$ corresponds to $M^2\approx1$, $Gr\approx900$.
Inset: peak value of the dimensional growth rate  (upper) and its location (lower) as functions of the domain size.}
\label{fig:corona1}
\end{figure}

\begin{figure}[ht]
\includegraphics[width=5in]{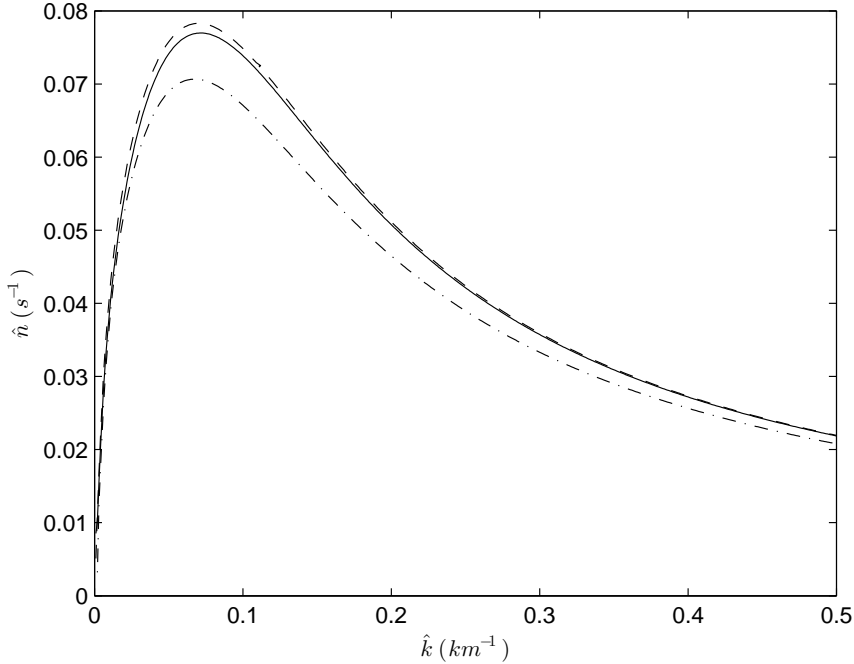}
\caption{The dimensional growth rates for the solar corona calculated in the  viscous-inviscid interface limit: $\hat{L}=200 km$ and $\hat{\mu}_1=10 kg/km s$ and $\hat{\mu}_2=0 kg/km s$. 
Solid line: constant temperature $\Theta=0$, $M^2\approx1$,
dashed line: $\Theta=-0.9$, $M^2\approx1$, $\xi=0$,
dashed-dot line: $\Theta=-0.9$, $M^2\approx1$, $\xi=2.5$.}
\label{fig:corona2}
\end{figure}

In Fig.~\ref{fig:corona1}, the growth rate is calculated based on the parameters derived from  \cite{Berger2010, Ballai2015, Delcroix1968} in the limit of the viscous-inviscid interface without temperature gradient. Similar to the previous calculations, the influence of the domain size is examined first. The inviscid-inviscid interface is also shown for comparison. The inset to Fig.~\ref{fig:corona1} demonstrates that, for $\hat{L} \gtrapprox 200km$ both the peak growth rate and its location almost do not change. As in the previous example, viscosity plays a significant role, especially at $\hat{k}>0.5 km^{-1}$.  Fig.~\ref{fig:corona2} shows the effect of $\Theta$ for the viscous-inviscid interface. Similar to the previous results, the effect is more pronounced for the temperature dependent viscosity, with the wavenumbers least affected by viscosity and showing significant growth lying in the range $\hat{k}\approx0.01 km^{-1}$ to $\hat{k}\approx 0.5 m^{-1}$.

\section{Conclusions}
\label{sec:6}

Using a normal mode analysis, the effects of viscosity and background temperature gradient, $\Theta$, on the growth rate are systematically studied for the compressible RTI with two immiscible fluids. When the effects of heat conduction are considered, a uniform $\Theta\neq 0$ still allows a steady background state and the linearized equations reduce to ordinary differential equations. This relaxes the assumption made in previous normal mode studies of compressible RTI, which satisfied the requirement of a steady background state by using uniform background temperature. Allowing $\Theta \neq 0$ makes the analysis closer to practical applications such as ICF and solar corona. The non-dimensional growth rate, $n^2/k=\hat{n}^2/\hat{k}\hat{g}$, is presented as a function of the non-dimensional wave number, $k=\hat{k}\hat{L}$, and analyzed for a range of $At$ and $Gr$ numbers, with positive and negative background temperature gradients, as well as  for compressible and incompressible flow ($p_\infty \rightarrow \infty$) cases.  The incompressible fluid limit \cite{Livescu2004,Yu2008,Livescu13} as well as the effects of different specific heats were not addressed since they are not directly relevant to the applications discussed in the paper. 

The results are presented as a logical set from three different interface types corresponding to combinations of the viscous properties of the two fluids: inviscid-inviscid, viscous-viscous and viscous-inviscid. The viscous-inviscid configuration has not been studied before in the context of RTI and provides a convenient way of addressing applications with large viscosity ratios between the two fluids, as in the examples mentioned above.

For two limiting cases, the inviscid-inviscid configuration at large $\Theta$ and  incompressible viscous-inviscid configuration with $\Theta=0$, the dispersion equations for the growth rate are obtained analytically. The former  case shows good correspondence with the numerical results at large $At$ and/or large wave numbers, so that the analytical result can be used instead of numerical calculations for such ranges of parameters. 

In general, for all cases, the effect of $\Theta<0$, corresponding to hotter light fluid, is found to be destabilizing and that of $\Theta>0$ stabilizing, compared to the background state with $\Theta=0$. These results are consistent with the corresponding background density stratifications. The effect of the $\Theta\neq 0$ is stronger at small $At$ and becomes small as $At$ approaches $1$ for all cases. In the limit of large $k$, the effect diminishes and the growth rates approach the corresponding $\Theta=0$ case. On the other hand, for the inviscid case, at small $k$ values, the growth rate obtained with $\Theta<0$ exceeds the infinite domain incompressible constant density result, $n^2/k=At$, so that this result is no longer an upper bound for the compressible growth rate as in the $\Theta=0$ case. Then, since $n^2/k$ corresponding to  $\Theta<0$ should approach the value $At$ from above, this suggest the existence of a peak in the normalized growth rate variation with $k$. The magnitude of the overshoot relative to the $n^2/k=At$ value decreases with $At$, consistent with the rest of the results. 

The effect of viscosity on the growth rate is important for all $At$ numbers and at all scales but becomes dominant at large $k$. As $Gr$ number increases, the viscous growth rates approach the limiting inviscid case results for some small range of $k$ and this range becomes larger with $At$. Viscous cases have a most unstable mode at $k\approx 2$ (the wavenumber location decreases with $At$) almost insensitive to $Gr$. The viscous-inviscid growth rate is closer to the fully inviscid case at high $At$, with larger differences at lower $At$ values. Compared to the viscous-viscous case, there are noticeable differences for the viscous-inviscid case results at all At values.

The numerical simulations are applied to two practical examples displaying  RTI -- coasting phase in ICF and solar corona plumes. The results demonstrate the importance of  inclusion of viscosity, which can significantly damp the growth rates starting from as small wave numbers as $\hat{k}\approx 0.5\mu m^{-1}$ for ICF and $\hat{k}\approx 0.02km^{-1}$ for solar corona. For both applications, $\Theta\neq 0$ has no significant influence on the growth rates for constant viscosities, but when viscosity is allowed to vary with temperature, the effect becomes noticeable.

\begin{acknowledgments}
The authors would like to thank H. Yu and R. McClarren for help with the early stages of this paper. 
This work was made possible in part by funding from the LDRD program at Los Alamos National Laboratory through project number 20150568ER. Los Alamos National Laboratory is operated by Los Alamos National Security, LLC for the US Department of Energy NNSA under Contract No. DE-AC52-06NA25396. Computational resources were provided by the LANL Institutional Computing (IC) Program.
\end{acknowledgments}

\appendix*

\section{Equation for the viscous case}

In the following derivations, the dynamic viscosity is assumed constant on the different sides of the interface  and the kinematic viscosity is considered continuous over interface, so that $\mu_{0,1}/\rho_{0,1}=\mu_{0,2}/\rho_{0,2}$, where $\mu_{0,m}/\rho_{0,m}=Gr^{-1/2}(\Theta \kappa_m M^2x_1+1)^{\alpha_m/(\Theta \kappa_m) +1}$.
The equations for $u_1$ and $\Delta$ on each side of the interface can be written as 
\begin{equation}\label{eq:collected_va}
A_4D^4u_1 + A_3D^3u_1 + A_2D^2u_1 + A_1Du_1 + A_0u_1 = 0 \:,
\end{equation}
\begin{equation}\label{eq:Delta_va}
\beta_1\Delta = B_3D^3u_1 + B_2D^2u_1 + B_1Du_1 + B_0u_1 \:,
\end{equation}
where the coefficients (with subscript $(_m)$ is dropped for simplicity) are given by
\begin{equation}\label{eq:A4_va}
A_4=\frac{\beta_2}{\alpha M^2n}B_3 \:,
\end{equation}
\begin{equation}\label{eq:A3_va}
A_3=\frac{\beta_2}{\alpha M^2n} \left( - \frac{D\beta_1B_3}{\beta_1}+DB_3+B_2 \right) -\frac{\gamma-1}{n}B_3 \:,
\end{equation}
\begin{equation}\label{eq:A2_va}
A_2=\frac{\beta_2}{\alpha M^2n} \left( - \frac{D\beta_1B_2}{\beta_1}+DB_2+B_1 \right) -\frac{\gamma-1}{n}B_2+\frac{\beta_3}{\alpha M^2n}\beta_1 \:,
\end{equation}
\begin{equation}\label{eq:A1_va}
A_1=\frac{\beta_2}{\alpha M^2n} \left( - \frac{D\beta_1B_1}{\beta_1}+DB_1+B_0 \right) -\frac{\gamma-1}{n}B_1-\frac{\beta_1}{n} \:,
\end{equation}
\begin{equation}\label{eq:A0_va}
A_0=\frac{\beta_2}{\alpha M^2n} \left( - \frac{D\beta_1B_0}{\beta_1}+DB_0 \right) -\frac{\gamma-1}{n}B_0- \left( n+\frac{k^2\beta_3}{\alpha M^2n} \right) \beta_1 \:,
\end{equation}
\begin{equation}\label{eq:B3_va}
B_3=-\frac{\beta_2\beta_3}{\alpha^3 M^6n^3} (\beta_2+\beta_3) \:,
\end{equation}
\begin{equation}\label{eq:B2_va}
B_2=\frac{\beta_3}{\alpha^2 M^4n^3} \left( \left( 1-\frac{D\beta_3}{\alpha M^2}\right) \beta_2-\left( \gamma -1-\frac{D\beta_2}{\alpha M^2} \right) \beta_3 \right) \:,
\end{equation}
\begin{equation}\label{eq:B1_va}
B_1=\frac{1}{\alpha M^2n^3} \left( \left( \gamma -1-\frac{D\beta_2}{\alpha M^2}\right) \beta_3+(\beta_2+\beta_3)\left( n^2+\frac{k^2\beta_3}{\alpha M^2}\right) \frac{\beta_2}{\alpha M^2}\right) \:,
\end{equation}
\begin{equation}\label{eq:B0_va}
B_0=\frac{1}{\alpha M^2n^3} \left( \left( \gamma -1-\frac{D\beta_2}{\alpha M^2}\right) \left( n^2+\frac{k^2\beta_3}{\alpha M^2}\right) \beta_3+\left( \frac{D\beta_3\beta_3}{\alpha M^2}+\beta_2\right) \frac{\beta_2k^2}{\alpha M^2}\right) \:,
\end{equation}
\begin{equation}\label{eq:beta3_va}
\beta_3=\frac{\alpha M^2n}{\sqrt{Gr}}(\Theta \kappa M^2x_1+1)^{\frac{\alpha}{\Theta \kappa }+1} \:,
\end{equation}
\begin{equation}\label{eq:beta2_va}
\beta_2=\gamma(\Theta \kappa M^2x_1+1)+\frac{1}{3}\beta_3 \:,
\end{equation}
\begin{equation}\label{eq:beta1_va}
\beta_1=\frac{1}{\alpha M^2n^3}\left( \left(n^2+\frac{k^2(\beta_2+\beta_3)}{\alpha M^2}\right) \beta_2^2-(\gamma-1)\beta_3\left( \gamma-1-\frac{D\beta_2}{\alpha M^2}\right) \right) \:,
\end{equation}
with $DB_3$, $DB_2$, $DB_1$, $DB_0$ obtained after differentiation of Eq. (\ref{eq:B3_va}) --  (\ref{eq:B0_va}), and $D\beta_3$, $D\beta_1$, $D\beta_1$ after differentiation of Eq. (\ref{eq:beta3_va}) --  (\ref{eq:beta1_va}), respectively (not shown). 

The equation for $D\Delta$ can be written as
\begin{equation}\label{eq:DDelta_va}
D\Delta=\frac{\alpha M^2n}{\beta_2}\left( \frac{\gamma-1}{n}\Delta-\frac{\beta_3}{\alpha M^2n}D^2u_1+\frac{1}{n}Du_1+\left( n+\frac{k^2\beta_3}{\alpha M^2n}\right) u_1 \right) \:,
\end{equation}
and the equation for $D^2\Delta$
\begin{multline}\label{eq:DDDelta_va}
D^2\Delta=\frac{\alpha M^2(\gamma-1)}{\beta_2}D\Delta-\frac{\alpha M^2D\beta_2(\gamma-1)}{\beta_2^2}\Delta-\frac{\beta_3}{\beta_2}D^3 u_1+\left( \frac{1}
{\beta_2^2}(D\beta_2\beta_3-\beta_2D\beta_3)+\frac{\alpha M^2}{\beta_2}\right) D^2u_1
\\+\frac{\alpha M^2}{\beta_2}\left( n^2+\frac{k^2\beta_3}{\alpha M^2}-\frac{D\beta_2}{\beta_2}\right) Du_1+\frac{1}{\beta_2^2}\left( k^2(\beta_2D\beta_3-D\beta_2\beta_3)-\alpha M^2n^2D\beta_2\right) u_1 \:.
\end{multline}

\bibliography{RT_temp_refs}

\end{document}